\documentclass[conference]{IEEEtran}

\usepackage{amsmath,amssymb,amsfonts}
\usepackage{bm}
\usepackage{booktabs}
\usepackage{graphicx}
\usepackage{subfig}
\usepackage{cite}
\usepackage[hidelinks]{hyperref}
\allowdisplaybreaks

\newcommand{\brho}{\boldsymbol{\rho}}
\newcommand{\bc}{\boldsymbol{c}}
\newcommand{\I}{\mathbb{I}}
\newcommand{\E}{\mathbb{E}}
\newcommand{\Tr}{\mathrm{Tr}}

\newcommand{\spn}{\mathrm{span}}

\begin{document}

\title{A Slow-Time Receiver Interface for Turbulent Free-Space Polarization Channels}

\author{%
\IEEEauthorblockN{Heyang Peng, Seid Koudia, and Symeon Chatzinotas}
\IEEEauthorblockA{Interdisciplinary Centre for Security, Reliability and Trust (SnT), University of Luxembourg, Luxembourg \\
Email: \{heyang.peng, seid.koudia, symeon.chatzinotas\}@uni.lu}
}

\maketitle

\begin{abstract}
The aperture-conditioned receiver interface introduced in the baseline model provides a compact description of turbulent free-space polarization channels, but it is static and therefore misses the temporal memory induced by atmospheric evolution. This paper develops a slow-time extension of that receiver-side interface. The hidden phase field, beam-centroid displacement, and scintillation are promoted to stochastic processes, and the interface is generated from their instantaneous realizations rather than from a prior ensemble average. A first-order local closure maps coarse-grained receiver-plane phase roughness to the variance of an effective polarization-mixing angle while preserving the inherited local polarization-channel family. After aperture conditioning, this yields time-dependent effective depolarization, coherence, and detection parameters. The resulting formulation remains compact, receiver-side, and protocol-facing: it exposes temporally correlated interface variables that can be block-aggregated in later quantum-communication performance or key-rate analyses, without requiring a full protocol-layer treatment in the present conference paper. Representative weak-turbulence results illustrate distinct fluctuation amplitudes and memory scales for the polarization and detection branches.
\end{abstract}


\section{Introduction}
\label{sec:intro}

Free-space optical (FSO) links are a natural platform for long-distance quantum communications because they avoid the exponential attenuation of long terrestrial fibers and support horizontal, airborne, and satellite deployments within a unified optical architecture \cite{ref_ursin,ref_yin,ref_takenaka,ref_pirandola_prr}. Their relevance extends from entanglement distribution and satellite quantum experiments to security-oriented protocols such as QKD and MDI-QKD \cite{ref_lo_mdi,ref_ghalaii_fso_mdi,ref_vasylyev_satellite}. In these links, atmospheric turbulence induces phase distortion, beam spreading, beam wander, and irradiance scintillation, which jointly determine how the receiver samples the optical field and how the detected quantum state should be represented \cite{ref_andrews,ref_churnside,ref_dios_uplink,ref_dios_temporal,ref_camboulives}.

Our baseline work introduced an aperture-conditioned receiver interface for turbulent free-space polarization channels \cite{ref_main}. That formulation reduces the received field to a finite-aperture spatial mixture of local polarization maps together with a scalar detection branch, yielding a compact receiver-side description in terms of effective depolarization, coherence loss, and collection probability. This interface is directly usable by higher-layer analysis, but it is static: the hidden atmospheric fluctuations are averaged out before the effective interface is formed.

The present paper asks a narrower question that is appropriate for a six-page conference contribution: how can the receiver-side interface itself be made time dependent in a physically structured way? The objective is not to append time labels to already averaged effective parameters, but to expose how a temporal receiver interface is generated from evolving hidden propagation quantities. To this end, we promote three hidden ingredients of the inherited interface to slow-time stochastic processes: a receiver-plane phase field, a centroid displacement process, and a scintillation process. A first-order local closure then converts coarse-grained receiver-plane phase roughness into the variance of an effective polarization-mixing angle, thereby preserving the inherited local channel family while inducing slow-time local polarization coefficients.

This paper makes three contributions. First, it formulates a \emph{generative} slow-time receiver interface in which hidden atmospheric realizations are retained prior to aperture conditioning. Second, it introduces a leading-order, calibration-ready closure linking local phase roughness to local polarization mixing, supported by standard spatiotemporal phase-statistics modeling and frozen-flow arguments \cite{ref_taylor1938,ref_vogel,ref_prasad}. Third, it identifies a compact temporal output,
\begin{equation}
\Xi_{\mathrm{red}}(t)
=
\bigl(
\lambda_a^{\mathrm{eff}}(t),
\,r_a^{2,\mathrm{eff}}(t),
\,\eta_{\mathrm{total}}(t)
\bigr),
\label{eq:Xi_intro}
\end{equation}
which is already protocol-facing: over slow-time blocks, these variables can be aggregated into receiver-side descriptors relevant to later FSO-QKD or MDI-QKD performance analysis \cite{ref_lo_mdi,ref_ghalaii_fso_mdi,ref_pirandola_prr,ref_klen_timecorr}.

The scope is intentionally limited. We assume a non-diattenuating receiver, a local slow-time observation window over which second-order statistics are approximately stationary, a leading-order scale separation between centroid motion and local phase roughness, and a radial numerical surrogate introduced only \emph{after} the exact two-dimensional interface is defined. Under these assumptions, the paper focuses on constructing and characterizing the temporal receiver interface itself, rather than on a full protocol-layer key-rate analysis.

\section{Slow-Time Receiver Interface Model}
\label{sec:model}

\subsection{Static receiver interface and scope}
\label{subsec:static_scope}

We retain the receiver-side philosophy of the baseline model \cite{ref_main}: the goal is not a full coherent propagation theory in which all spatial modes are preserved to the end, but a reduced interface that is directly consumable by communication-layer analysis. Let
\begin{equation}
\mathcal{H}_P=\spn\{|H\rangle,|V\rangle\}
\end{equation}
denote the polarization Hilbert space, and let the input polarization qubit be
\begin{equation}
\hat{\rho}_P=
\begin{pmatrix}
\rho_{HH} & \rho_{HV}\\
\rho_{VH} & \rho_{VV}
\end{pmatrix},
\qquad
\hat{\rho}_P\ge 0,
\qquad
\Tr\hat{\rho}_P=1.
\end{equation}
At the receiver plane, the transverse coordinate is $\brho=(x,y)\in\mathbb{R}^2$, the aperture is the disk $A=\{\brho:\|\brho\|\le a\}$, and the aperture projector is
\begin{equation}
\Pi_a=\int_A |\brho\rangle\langle\brho|\,d^2\brho.
\end{equation}
If $\rho_{SP}$ is the reduced spatial--polarization state before aperture conditioning, then the detected polarization state and aperture collection probability are
\begin{equation}
\hat{\rho}'_P=
\frac{\Tr_S[\Pi_a\rho_{SP}\Pi_a]}{\Tr[\Pi_a\rho_{SP}\Pi_a]},
\qquad
\eta_{\mathrm{eff}}=\Tr[\Pi_a\rho_{SP}\Pi_a].
\label{eq:basic_detected_state}
\end{equation}
Under the present receiver model, attenuation and scintillation are polarization-insensitive scalar factors, so
\begin{equation}
\eta_{\mathrm{total}}=\eta_0 I_{\mathrm{scin}}\eta_{\mathrm{eff}},
\label{eq:eta_total_static}
\end{equation}
where $\eta_0$ collects deterministic efficiencies. Because the receiver is assumed non-diattenuating, such scalar factors do not modify the normalized state $\hat{\rho}'_P$.

At the receiver-interface level, the static baseline represents the spatial degree of freedom as an aperture-conditioned classical mixture of local polarization maps,
\begin{equation}
\hat{\rho}'_P=
\frac{\E_{\brho_d}\!\left[\int_A \omega(\brho,\brho_d)\,\mathcal{E}_{\mathrm{phase},\brho}(\hat{\rho}_P)\,d^2\brho\right]}
{\E_{\brho_d}\!\left[\int_A \omega(\brho,\brho_d)\,d^2\brho\right]},
\label{eq:static_mixture}
\end{equation}
with
\begin{equation}
\eta_{\mathrm{eff}}=
\E_{\brho_d}\!\left[\int_A \omega(\brho,\brho_d)\,d^2\brho\right].
\label{eq:static_etaeff}
\end{equation}
Here, $\omega(\brho,\brho_d)$ is the receiver-plane spatial weight induced by beam spreading and beam drift, and $\mathcal{E}_{\mathrm{phase},\brho}$ is the local polarization degradation map. The present paper preserves this architecture and only changes the order of operations: hidden atmospheric realizations are retained first, and temporal statistics are formed afterwards.

\subsection{Hidden processes and first-order closure}
\label{subsec:hidden_closure}

The inherited interface depends on three physical ingredients: phase-induced local polarization degradation, centroid-driven aperture sampling, and scintillation-driven scalar intensity fluctuation. We therefore introduce the hidden slow-time state
\begin{equation}
\mathcal{Z}(t)=\bigl(\phi(\brho,t),\,\bc(t),\,I_{\mathrm{scin}}(t)\bigr),
\label{eq:hidden_state}
\end{equation}
where $\phi(\brho,t)$ is the receiver-plane phase field, $\bc(t)$ is the beam-centroid displacement process, and $I_{\mathrm{scin}}(t)$ is the scintillation process.

Although $\phi$ and $\bc$ originate from the same turbulent medium, they enter the receiver interface through different effective mechanisms. Beam wander is primarily associated with large-scale, low-spatial-frequency refractive fluctuations \cite{ref_churnside,ref_yura,ref_dios_uplink}, whereas local phase roughness is associated with finer spatial phase variations and their receiver-side reduction \cite{ref_andrews,ref_vogel,ref_prasad}. We therefore adopt a \emph{leading-order} scale-separated interface model. This is an interface approximation, not a claim of exact physical independence: residual couplings may be reintroduced through the joint law of $(\phi,\bc,I_{\mathrm{scin}})$ without changing the receiver-side architecture.

The variable $t$ denotes \emph{channel slow time}. It is not optical carrier time, not wave-propagation time, and not an additional quantum degree of freedom. It indexes the slowly evolving atmospheric realization observed across successive channel uses. Over a local observation window, we assume approximate wide-sense stationarity of second-order statistics.

Let $\delta n(\brho,\xi,t)$ be the refractive-index fluctuation along the propagation path, where $\xi$ is the longitudinal coordinate and $k=2\pi/\lambda$ is the optical wavenumber. The accumulated phase is
\begin{equation}
\phi(\brho,z,t)=k\int_0^z \delta n(\brho,\xi,t)\,d\xi,
\qquad
\phi(\brho,t):=\phi(\brho,z_{\mathrm{rx}},t).
\end{equation}
Its second-order statistics may be expressed through the covariance and structure function,
\begin{equation}
C_\phi(\Delta\brho,\tau)=\E\!\left[\phi(\brho,t)\phi(\brho+\Delta\brho,t+\tau)\right],
\end{equation}
\begin{equation}
D_\phi(\Delta\brho,\tau)=\E\!\left[\bigl(\phi(\brho+\Delta\brho,t+\tau)-\phi(\brho,t)\bigr)^2\right].
\end{equation}
To instantiate temporal dependence, we adopt a frozen-flow picture,
\begin{equation}
\phi(\brho,t+\tau)\approx \phi(\brho-\mathbf{v}_\perp\tau,t),
\end{equation}
which traces back to Taylor's hypothesis and is routinely used in optical phase modeling \cite{ref_taylor1938,ref_prasad}. A path-integrated Kolmogorov-type form is then
\begin{equation}
D_\phi(\Delta\brho,\tau)=2.91\,k^2\int_0^{z_{\mathrm{rx}}} C_n^2(\xi)
\left\|\Delta\brho-\mathbf{v}_\perp(\xi)\tau\right\|^{5/3} d\xi.
\label{eq:Dphi_path}
\end{equation}

The instantaneous interface, however, requires a realization-level local quantity at each $(\brho,t)$, not only ensemble statistics. We therefore introduce a coarse-grained local phase roughness
\begin{equation}
\mathcal{J}_\ell(\brho,t)=\int_{\mathbb{R}^2} K_\ell(\Delta\brho)
\bigl[\phi(\brho+\Delta\brho,t)-\phi(\brho,t)\bigr]^2 d^2\Delta\brho,
\label{eq:J_roughness}
\end{equation}
where $K_\ell\ge 0$ is a normalized kernel with spatial scale $\ell$. This definition is consistent with the standard structure-function picture because
\begin{equation}
\E[\mathcal{J}_\ell(\brho,t)]=\int_{\mathbb{R}^2} K_\ell(\Delta\brho)D_\phi(\Delta\brho,0)\,d^2\Delta\brho.
\label{eq:EJ}
\end{equation}

The mapping from local phase roughness to an effective polarization-mixing variance is introduced here as a receiver-side closure, not as a microscopic identity. Its role is to preserve the inherited local channel family while exposing slow-time dependence through a small number of interpretable interface parameters. We write
\begin{equation}
\sigma_\theta^2(\brho,t)=F\!\left(\mathcal{J}_\ell(\brho,t)\right),
\qquad F(0)=0,
\end{equation}
and in the weak local-perturbation regime retain the leading term,
\begin{equation}
\sigma_\theta^2(\brho,t)\approx \gamma_\theta\mathcal{J}_\ell(\brho,t).
\label{eq:sigma_closure}
\end{equation}
Here, $\ell$ sets the receiver-side reduction scale and $\gamma_\theta$ sets the closure sensitivity. In this conference treatment they are scenario-level interface parameters, not time-varying fitting knobs.

For small coarse-graining scale, a local gradient expansion gives
\begin{equation}
\phi(\brho+\Delta\brho,t)-\phi(\brho,t)\approx \Delta\brho\cdot\nabla_\perp\phi(\brho,t),
\end{equation}
so that, for an isotropic Gaussian kernel,
\begin{equation}
K_\ell(\Delta\brho)=\frac{1}{\pi\ell^2}\exp\!\left(-\frac{\|\Delta\brho\|^2}{\ell^2}\right),
\end{equation}
we obtain the scale interpretation
\begin{equation}
\mathcal{J}_\ell(\brho,t)\approx \frac{\ell^2}{2}\,\|\nabla_\perp\phi(\brho,t)\|^2,
\qquad
\gamma_\theta\propto \frac{1}{(k\ell)^2}.
\label{eq:gamma_scale}
\end{equation}
This aligns the closure with coarse-grained wavefront-slope energy and with standard spatiotemporal covariance modeling \cite{ref_vogel,ref_prasad}.

Once $\sigma_\theta^2(\brho,t)$ is available, the temporal local polarization coefficients follow from the inherited local channel family \cite{ref_main}. Assuming a conditional Gaussian law for the effective local mixing angle,
\begin{equation}
p_\theta(\theta|\brho,t)=\frac{1}{\sqrt{2\pi\sigma_\theta^2(\brho,t)}}
\exp\!\left[-\frac{\theta^2}{2\sigma_\theta^2(\brho,t)}\right],
\end{equation}
we have $\E[e^{i\theta}|\brho,t]=\exp[-\sigma_\theta^2(\brho,t)/2]$, and therefore
\begin{equation}
\lambda(\brho,t)=\frac{1-\exp[-\sigma_\theta^2(\brho,t)/2]}{2},
\label{eq:lambda_local}
\end{equation}
\begin{equation}
\bar r^2(\brho,t)=1-\lambda(\brho,t)\frac{3\mu_\parallel(\kappa)-1}{2}.
\label{eq:rbar_local}
\end{equation}
These are the slow-time counterparts of the local depolarization and coherence descriptors of the baseline interface.

\subsection{Instantaneous interface and radial surrogate}
\label{subsec:instantaneous}

At slow-time instant $t$, the local polarization map is
\begin{equation}
\mathcal{E}_{\mathrm{phase},\brho,t}(\hat\rho_P)=
(1-\lambda(\brho,t))
\begin{pmatrix}
\rho_{HH} & \rho_{HV}\bar r^2(\brho,t)\\
\rho_{VH}\bar r^2(\brho,t) & \rho_{VV}
\end{pmatrix}
+\frac{\lambda(\brho,t)}{2}\I.
\label{eq:local_map_time}
\end{equation}
Define also
\begin{equation}
q(\brho,t):=(1-\lambda(\brho,t))\bar r^2(\brho,t),
\label{eq:q_local}
\end{equation}
which will control the local action on the coherence terms.

At the spatial branch, the centroid realization $\bc(t)$ sets the instantaneous receiver-plane weight. Retaining the Gaussian-beam form of the baseline model,
\begin{equation}
\omega(\brho,t)=\frac{2}{\pi w_z^2}
\exp\!\left(-\frac{2\|\brho-\bc(t)\|^2}{w_z^2}\right),
\label{eq:omega_time}
\end{equation}
where $w_z$ is the effective receiver-plane beam radius within the local observation window. The time-resolved receiver-plane state is then represented as the classical mixture
\begin{equation}
\rho_{SP}(t)=\int_{\mathbb{R}^2}\omega(\brho,t)
|\brho\rangle\langle\brho|\otimes \mathcal{E}_{\mathrm{phase},\brho,t}(\hat\rho_P)\,d^2\brho.
\label{eq:rhoSP_time}
\end{equation}
Applying aperture conditioning yields
\begin{equation}
\hat\rho'_P(t)=
\frac{\int_A \omega(\brho,t)\mathcal{E}_{\mathrm{phase},\brho,t}(\hat\rho_P)\,d^2\brho}
{\int_A \omega(\brho,t)\,d^2\brho},
\qquad
\eta_{\mathrm{eff}}(t)=\int_A \omega(\brho,t)\,d^2\brho.
\label{eq:instantaneous_interface}
\end{equation}
Substituting \eqref{eq:local_map_time} into \eqref{eq:instantaneous_interface} and integrating entrywise give
\begin{equation}
\hat\rho'_P(t)=
\begin{pmatrix}
(1-\Lambda_a(t))\rho_{HH}+\Lambda_a(t)/2 & Q_a(t)\rho_{HV}\\
Q_a(t)\rho_{VH} & (1-\Lambda_a(t))\rho_{VV}+\Lambda_a(t)/2
\end{pmatrix},
\label{eq:rhoP_reduced_time}
\end{equation}
with
\begin{equation}
\Lambda_a(t)=
\frac{\int_A \omega(\brho,t)\lambda(\brho,t)\,d^2\brho}
{\int_A \omega(\brho,t)\,d^2\brho},
\qquad
Q_a(t)=
\frac{\int_A \omega(\brho,t)q(\brho,t)\,d^2\brho}
{\int_A \omega(\brho,t)\,d^2\brho}.
\label{eq:Lambda_Q_time}
\end{equation}
The exact aperture-conditioned output is therefore the pair $(\Lambda_a(t),Q_a(t))$, rather than $\bar r^2$ alone. The effective temporal parameters are defined by
\begin{equation}
\lambda_a^{\mathrm{eff}}(t):=\Lambda_a(t),
\qquad
r_a^{2,\mathrm{eff}}(t):=\frac{Q_a(t)}{1-\Lambda_a(t)},
\label{eq:effective_time_params}
\end{equation}
so that
\begin{equation}
\hat\rho'_P(t)=
(1-\lambda_a^{\mathrm{eff}}(t))
\begin{pmatrix}
\rho_{HH} & \rho_{HV}r_a^{2,\mathrm{eff}}(t)\\
\rho_{VH}r_a^{2,\mathrm{eff}}(t) & \rho_{VV}
\end{pmatrix}
+\frac{\lambda_a^{\mathrm{eff}}(t)}{2}\I.
\label{eq:rhoP_baseform_time}
\end{equation}
Under the present receiver assumptions,
\begin{equation}
\eta_{\mathrm{total}}(t)=\eta_0I_{\mathrm{scin}}(t)\eta_{\mathrm{eff}}(t),
\label{eq:eta_total_time}
\end{equation}
and $I_{\mathrm{scin}}(t)$ cancels out of the normalized state construction.

The two-dimensional formulation above is the exact receiver-interface definition adopted in this paper. For numerical evaluation, we then introduce a \emph{radial surrogate}. When aperture-integrated observables are dominated by centroid-induced anisotropy and the remaining high-spatial-frequency angular roughness is effectively averaged by the finite aperture, one may replace
\begin{equation}
\lambda(\brho,t)\rightarrow \lambda(r,t),
\qquad
q(\brho,t)\rightarrow q(r,t),
\qquad r=\|\brho\|.
\label{eq:radial_surrogate}
\end{equation}
Let $c(t):=\|\bc(t)\|$ and align the polar axis with $\bc(t)$. Then
\begin{equation}
\omega(r,\varphi,t)=\frac{2}{\pi w_z^2}
\exp\!\left[-\frac{2(r^2+c^2(t)-2rc(t)\cos\varphi)}{w_z^2}\right].
\end{equation}
Defining the azimuthally integrated radial weight
\begin{equation}
W(r,t):=\int_0^{2\pi}\omega(r,\varphi,t)\,r\,d\varphi,
\end{equation}
and using $\int_0^{2\pi}e^{\xi\cos\varphi}d\varphi=2\pi I_0(\xi)$ gives
\begin{equation}
W(r,t)=\frac{4r}{w_z^2}
\exp\!\left[-\frac{2(r^2+c^2(t))}{w_z^2}\right]
I_0\!\left(\frac{4rc(t)}{w_z^2}\right).
\label{eq:Wr}
\end{equation}
Hence,
\begin{equation}
\eta_{\mathrm{eff}}(t)=\int_0^a W(r,t)\,dr,
\end{equation}
\begin{equation}
\Lambda_a(t)=\frac{\int_0^a W(r,t)\lambda(r,t)\,dr}{\int_0^a W(r,t)\,dr},
\qquad
Q_a(t)=\frac{\int_0^a W(r,t)q(r,t)\,dr}{\int_0^a W(r,t)\,dr}.
\label{eq:radial_reduced_quantities}
\end{equation}
This surrogate is introduced only after the exact interface is defined; outside its validity regime, the full two-dimensional integrals should be retained.

\section{Protocol-Facing Temporal Statistics}
\label{sec:stats}

The slow-time extension yields a receiver-side stochastic process rather than a static channel triple. The purpose of this section is therefore twofold: to characterize temporal memory at the interface level, and to show why the resulting variables are already relevant to quantum-communication analysis.

For any scalar component $X(t)$ of the reduced interface, define the autocovariance and normalized autocorrelation
\begin{equation}
R_X(\tau)=\E\!\left[(X(t)-\mu_X)(X(t+\tau)-\mu_X)\right],
\qquad
\rho_X(\tau)=\frac{R_X(\tau)}{R_X(0)},
\label{eq:autocov_autocorr}
\end{equation}
and the correlation time
\begin{equation}
\tau_c^{(X)}=\frac{1}{R_X(0)}\int_0^{\infty}R_X(\tau)\,d\tau.
\label{eq:corr_time}
\end{equation}
For representative pairs $(X,Y)$ we also use the lagged cross-correlation
\begin{equation}
C_{XY}(\tau)=
\frac{\E\!\left[(X(t)-\mu_X)(Y(t+\tau)-\mu_Y)\right]}{\sigma_X\sigma_Y}.
\label{eq:crosscorr}
\end{equation}

These statistics are not introduced only for channel phenomenology. In FSO-QKD and MDI-QKD, protocol performance depends on temporally varying transmission, detection probability, and state quality rather than on a single static channel value \cite{ref_lo_mdi,ref_ghalaii_fso_mdi,ref_pirandola_prr,ref_vasylyev_weakstrong,ref_klen_timecorr}. The present interface does not yet produce a key rate; instead, it provides the temporally resolved receiver-side variables from which blockwise descriptors can later be formed. For a slow-time block $\mathcal{B}_m$, one may define, for example,
\begin{equation}
\bar\eta_m=\frac{1}{|\mathcal{B}_m|}\sum_{t\in\mathcal{B}_m}\eta_{\mathrm{total}}(t),
\qquad
\bar Q_m=\frac{1}{|\mathcal{B}_m|}\sum_{t\in\mathcal{B}_m}Q_a(t),
\label{eq:block_descriptors}
\end{equation}
together with the corresponding block-averaged polarization descriptors. These quantities constitute a receiver-side bridge between the temporal channel model developed here and later blockwise performance, estimation, or key-rate analyses.

\section{Representative Results}
\label{sec:results}

We present representative weak-turbulence results to verify that the proposed slow-time construction produces a numerically stable and physically interpretable receiver interface. The objective is not an exhaustive regime sweep in the conference version, but a focused validation of two central claims of the model: first, that the polarization and detection branches fluctuate with different amplitudes; and second, that they generally exhibit different temporal memory scales.

\begin{figure}[t]
    \centering
    \IfFileExists{Temporal_Interface_Traces_Weak.pdf}{%
        \includegraphics[width=\linewidth]{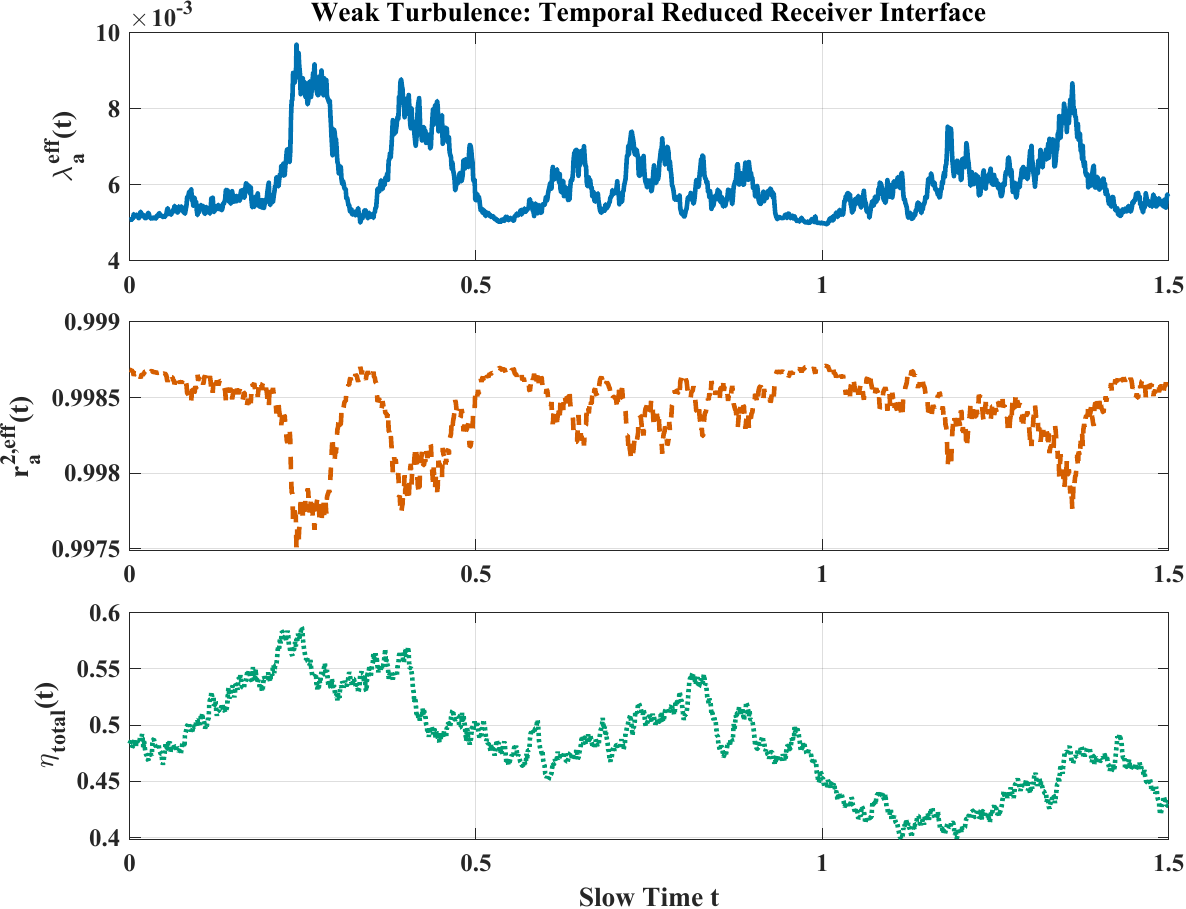}%
    }{%
        \fbox{\parbox[c][5.0cm][c]{0.95\linewidth}{\centering Placeholder for temporal-trace figure\\(insert \texttt{Temporal\_Interface\_Traces\_Weak.pdf})}}%
    }
    \caption{Representative weak-turbulence temporal traces of $\lambda_a^{\mathrm{eff}}(t)$, $r_a^{2,\mathrm{eff}}(t)$, and $\eta_{\mathrm{total}}(t)$. The polarization branch stays close to the near-ideal regime, while the scalar detection branch exhibits more pronounced slow-time fluctuation.}
    \label{fig:traces}
\end{figure}

Figure~\ref{fig:traces} shows representative traces of $\lambda_a^{\mathrm{eff}}(t)$, $r_a^{2,\mathrm{eff}}(t)$, and $\eta_{\mathrm{total}}(t)$ under weak turbulence. The effective depolarization remains small and the effective coherence remains close to unity, indicating that the polarization branch stays near the ideal regime after aperture-conditioned averaging. By contrast, the scalar detection branch exhibits visibly stronger fluctuations. This separation is physically meaningful: weak turbulence perturbs the normalized detected polarization state only mildly, while the scalar branch remains directly sensitive to centroid-driven aperture sampling and irradiance fluctuation \cite{ref_dios_uplink,ref_dios_temporal,ref_camboulives}.

\begin{figure}[t]
    \centering
    \IfFileExists{Temporal_Correlation_Weak.pdf}{%
        \includegraphics[width=\linewidth]{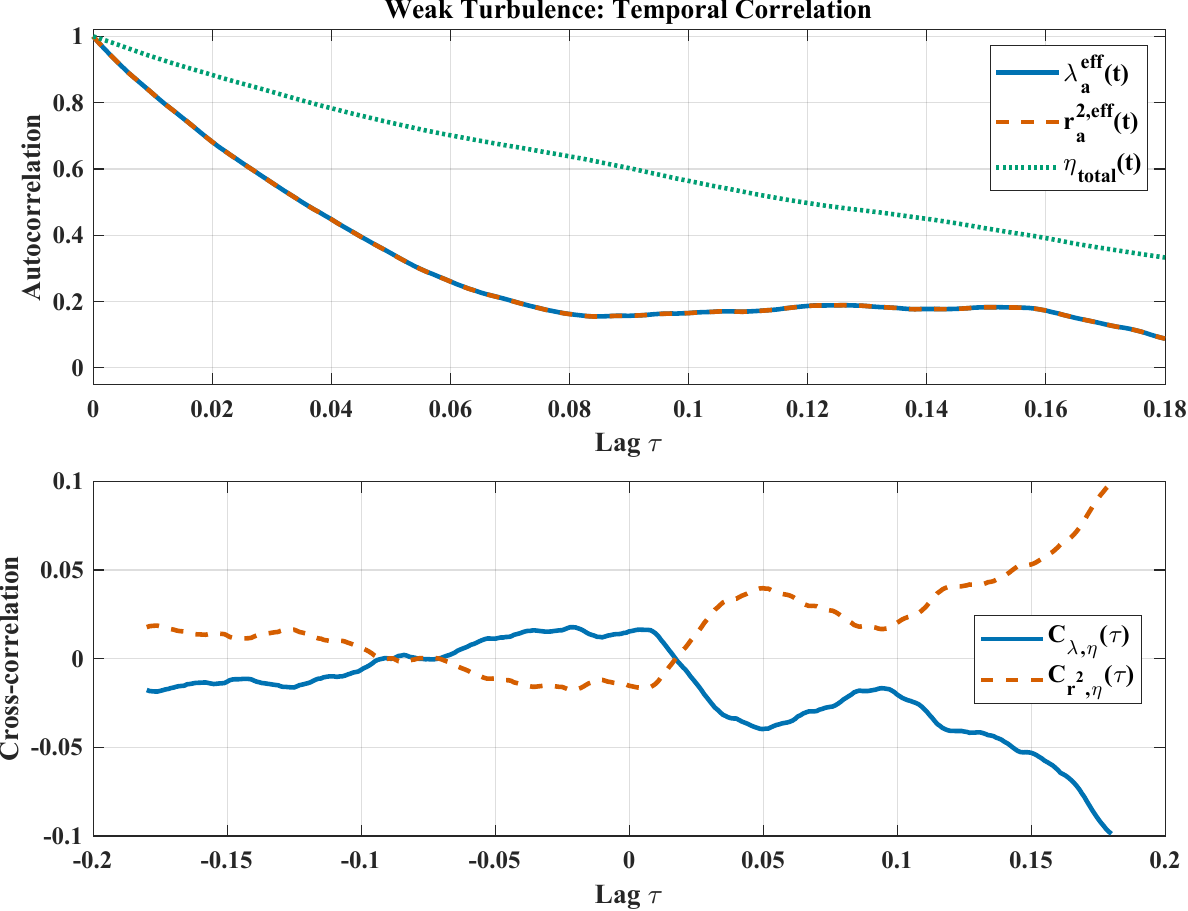}%
    }{%
        \fbox{\parbox[c][5.0cm][c]{0.95\linewidth}{\centering Placeholder for temporal-correlation figure\\(insert \texttt{Temporal\_Correlation\_Weak.pdf})}}%
    }
    \caption{Weak-turbulence temporal correlation structure of the reduced receiver interface. The upper panel shows the autocorrelation functions of $\lambda_a^{\mathrm{eff}}(t)$, $r_a^{2,\mathrm{eff}}(t)$, and $\eta_{\mathrm{total}}(t)$; the lower panel shows representative lagged cross-correlations between the polarization and detection branches.}
    \label{fig:corr}
\end{figure}

Figure~\ref{fig:corr} shows the corresponding temporal correlations. The autocorrelation curves of $\lambda_a^{\mathrm{eff}}(t)$ and $r_a^{2,\mathrm{eff}}(t)$ nearly overlap, which is consistent with their common origin in the phase-roughness branch and the shared aperture weighting. The autocorrelation of $\eta_{\mathrm{total}}(t)$ decays more slowly, indicating a longer memory in the detection branch. The lagged cross-correlations between polarization and detection variables remain nonzero but modest, which supports the leading-order branch decomposition without suggesting complete independence. This is precisely the kind of structured temporal information that a static interface cannot provide.

Although the present conference paper stops at the receiver-interface level, the output variables are already usable as blockwise inputs for later protocol studies. In particular, the unequal memory scales of polarization and detection branches imply that a single static parameter is generally insufficient to represent the channel over time windows relevant to estimation, postselection, or key-rate aggregation \cite{ref_pirandola_prr,ref_ghalaii_fso_mdi,ref_klen_timecorr}.

\section{Conclusion}
\label{sec:conclusion}

This paper developed a slow-time extension of an aperture-conditioned receiver interface for turbulent free-space polarization channels. Instead of averaging hidden atmospheric effects before reduction, the receiver-plane phase field, centroid displacement, and scintillation were retained as slow-time stochastic processes and mapped to an instantaneous receiver interface. A leading-order closure linked coarse-grained local phase roughness to effective polarization mixing, preserving the inherited local channel family while generating time-dependent local polarization coefficients. After aperture conditioning, this yielded a compact temporal receiver-side description in terms of effective depolarization, coherence, and total detection probability.

The contribution of the conference paper is therefore not a full protocol-level security analysis, but a protocol-facing temporal channel interface. It exposes the slow-time variables and memory measures that later blockwise FSO-QKD or MDI-QKD analyses can consume, while keeping the present treatment compact enough for the conference setting. Natural next steps are to extend the numerical study beyond weak turbulence and to connect the temporal interface to blockwise protocol evaluation.

\end{document}